\documentclass[article,twocolumn,author-numerical,showpacs]{revtex4-1}

\usepackage{amsmath,amssymb}
\usepackage{graphicx}
\usepackage{dcolumn}
\usepackage{bm}
 \usepackage{epsfig}
 \usepackage{color,soul}

\begin{document}

\title{Entrainment and Unit Velocity: Surprises in an Accelerated Exclusion
Process}
\author{Jiajia Dong$^{1,2}$, Stefan Klumpp$^{2}$, and Royce K.P. Zia$^{3}$}
\address{$^{1}$ Department of Physics and Astronomy, Bucknell University, Lewisburg, PA 17837\\
$^{2}$ Max Planck Institute of Colloids and Interfaces, 14424 Potsdam, Germany \\
$^{3}$ Physics Department, Virginia Tech, Blacksburg, VA, 24061 and \\
Department of Physics and Astronomy, Iowa State University, Ames, Iowa
50011}
\date{\today}

\begin{abstract}
We introduce a class of distance-dependent interactions in an accelerated exclusion process (AEP) inspired by the observation of transcribing RNA polymerase speeding up when ``pushed'' by a trailing one. On a ring, the AEP steady state displays a discontinuous transition, from being homogeneous (with augmented currents) to phase-segregated. In the latter state, the holes appear loosely bound and move together, much like a train. Surprisingly, the current-density relation is simply $J=1-\rho$, signifying that the ``hole-train'' travels with \textit{unit} velocity.
\end{abstract}


\maketitle
\begin{figure}[bph]
\vspace{-0.5cm}
\begin{center}
\includegraphics[height=1.7cm,width=9cm]{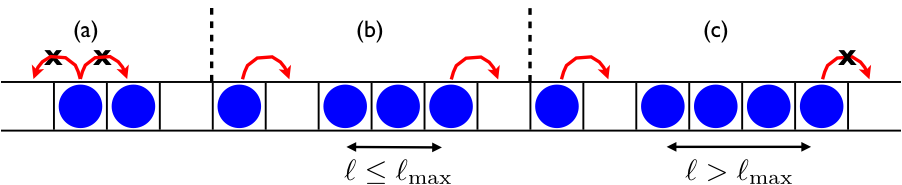}
\end{center}
\vspace{-0.6cm}
\caption{AEP with $\ell _{\max }=3$. (a) Backward hops and overlaps are forbidden. (b) A particle hopping to the back of a
3-cluster triggers an additional hop. No other hops are triggered, e.g., (c): particle hits a cluster with $%
\ell >\ell _{\max }$.}
\label{fig:diagram}
\end{figure}

Over four decades ago, the totally asymmetric simple exclusion process
(TASEP) was introduced by two distinct communities:
biochemistry and pure mathematics \cite{MGP68,MG69,Spitzer70}.
This venerable model has since enjoyed much attention, especially from
statistical physicists \cite{Spohn91,Schutz00}. Variations to
the original TASEP emerged, as different features are recognized to be
crucial for capturing essential aspects in biological and physical systems. For
example, for applications to various transport phenomena in molecular
biology \cite{Lipowsky06,CMZ11}, these additions include extended objects and
inhomogeneous hopping \cite{MGP68,MG69,SKZ03,LC03,DSZ07a,DSZ07b},
non-conserving particle numbers and multiple species \cite{EFGM95, GreulichS09}, ``recycling'' and competition \cite{CHOU2003,COOK2009A,COOK2009B,CZ2012,GCAR2012}. Inspired by the cooperative increase in speed in transcribing RNA polymerases (RNAP) that ``push'' each other \cite{Epshtein2003,Jin2010}, we set out to study an extension, in which the particle arriving at the rear of a cluster of particles ``triggers'' the particle at the front to move (Fig.\ref{fig:diagram}). A mechanical ``push'' example is Newton's balls. In general, the mechanism of ``pushing'' need not to be mechanical or involve actual collisions.

While interacting driven lattice gases have been studied for nearly
30 years \cite{KLS84,SZ95}, such ``facilitated'' action was considered only recently \cite
{AS00,BM09,GKR10,GR11}. Our model differs substantially
from these previous studies:
In \cite{AS00}, neighboring particles may attract/repel each other or a particle
may move as far as two sites, but do not trigger others at a distance to
move. Implied by the original name -- \textit{restricted} asymmetric
exclusion process \cite{BM09}, particles (with non-zero headway) in the
``facilitated asymmetric exclusion'' process \cite{GKR10} do not move unless there is one or more particles
behind it. In the more general ``cooperative
exclusion process'' \cite{GR11}, these rules are used a
fraction of the time while ordinary TASEP rules apply otherwise. Thus, it is
natural that the average current $J$ (as a function of $\rho $, the particle density on a ring) is always lower than the standard $J_{\text{TASEP}}=\rho \left( 1-\rho \right) $. By contrast, in our model  
$J$ is \textit{always higher}. Thus, we name our model the ``accelerated
exclusion process'' (AEP). Beyond this expected increase in $J$, we discover a
qualitatively new phenomenon. Typically, there are \textit{two branches} of 
$J\left( \rho \right) $: An augmented current (AC) branch at lower densities,
and a branch of lower current at high densities, with a \textit{discontinuous}
jump as $\rho $ crosses some critical value. Even more remarkable is that $%
J=1-\rho $ in the latter branch, i.e., the system has \textit{unit
velocity} (UV) on the average -- with particles and holes in opposite
directions. Such a simple property emerges from a system with
interacting particles is astonishing; it behooves us to name this state
the ``UV phase.'' Preliminary studies
indicate that its presence depends on two competing factors, to be
detailed below. In this letter, we first provide the biological motivation and define our model. Simulation
results, as well as some analytic understanding, will follow. In a final
paragraph, we outline the issues which should be pursued more
systematically.
\vspace{-0.5cm}

\textit{Motivations and Model}:  Our model is motivated by a cooperative effect in transcription where the forward motion of an RNAP is accelerated by the presence of a trailing one that prevents the first RNAP from entering alternative kinetic pathways such as pausing and backtracking  \cite{Epshtein2003,Jin2010}. Although the pausing and backtracking aspects were incorporated in an exclusion process earlier \cite{Klumpp2011}, facilitated motion was not studied\cite{footnote}. Another instance of ÓpushingÓ in transport is ÓtailgatingÓ in ve- hicular and pedestrian traffic, where the acceleration is mediated by the transfer of information (rather than mo- mentum as in NewtonÕs balls). Here we model the generic cooperative aspect of ÓpushingÓ with the following rules: Consider a 1-D lattice with \textit{periodic} boundary conditions -- for simplicity and as a baseline study. Let the occupation variable
of site $i$ ($=1,...,L$) be $n_{i}=1$ or $0$, corresponding to
a particle or a hole. In standard TASEP, the system evolves by
discrete attempt steps: Picking a random occupied site and moving the
particle to the next site provided the target is empty. The total number of
particles ($N\equiv \Sigma _{i}n_{i}$) is a constant, and the
particle density ($\rho \equiv N/L$) acts as a control parameter. When the
system settles down into a steady state, one of the non-trivial quantities of interest is the
average current, $J$ (average number of particle hops in an attempt) and its
dependence on $\rho $. Clearly, $J_{\text{TASEP}}=\left\langle n_{i}\left(
1-n_{i+1}\right) \right\rangle $, where the average is over all possible
configurations $\left\{ n_{i}\right\} $ with the appropriate weight. Since
Spitzer \cite{Spitzer70} showed that all $\left\{ n_{i}\right\} $ (with a fixed $N$) in the
steady state are equally probable, finding an exact expression for $J\left(
\rho ;L\right) $ is simple \cite{SKZ03} and $J_{\text{TASEP}}=\rho \left( 1-\rho
\right) $ emerges in the thermodynamic limit.

\begin{figure}[bph]
\vspace{-0.5cm}
\begin{center}
\includegraphics[height=4cm,width=9cm]{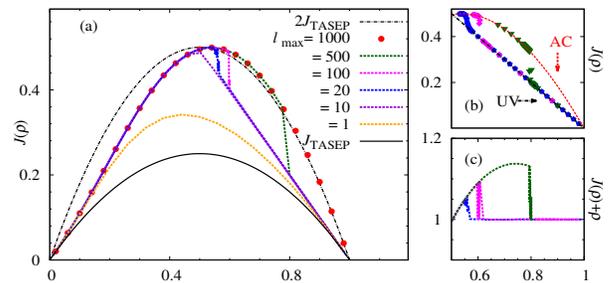}
\end{center}
\vspace{-0.7cm}
\caption{(a) $J_\text{AEP}(\rho)$ for various $\ell_{\max}$'s.
Detailed views of $\ell _{\max }=20,100,500$: symbols in (b), with dashed lines for $J_\text{AC}$ and $J_\text{UV}$ and in (c), $J(\rho)+\rho$ is plotted to accentuate the transition into UV.}
\label{fig:J_rho}
\end{figure}

For AEP, we include an additional move to account for long-range
interactions: If a particle moves to a vacant site and becomes the rear
of a particle cluster, it ``triggers'' the particle at the front of the cluster to move in the \textit{same} attempt. 
Note that the second particle does \textit{not}
trigger another move, even if it arrives at the back of another cluster.
Since particles are indistinguishable, we can also regard this action as the
chosen particle being ``accelerated''
through the entire cluster it hits! An exact expression for the current is 
$J_{\text{AEP}}=\left\langle n_{i}\left( 1-n_{i+1}\right) \right\rangle
+\left\langle n_{i}\left( 1-n_{i+1}\right) n_{i+2}\right\rangle $. Being
typically \textit{higher} than $J_{\text{TASEP}}$, the term ``augmented current'' (AC) seems apt. It is reasonable
to consider finite range for such interaction, an aspect we implement by
allowing only clusters of length up to $\ell _{\max }$ to facilitate the
accelerated move summarized in Fig.\ref{fig:diagram}. This length scale is dependent of the specific system at hand and more realistic rules can be introduced, e.g., the rate of acceleration being a smooth function of $\ell $ rather than the step function. Although $J_\text{AEP}\left( \rho \right)  >J_\text{TASEP}\left( \rho \right) $ is still expected, simulations reveal the
existence of AC and UV branches, with a discontinuous transition when a critical density, $\rho _{c}\left( \ell _{\max }\right) $,
is crossed (Fig.\ref{fig:J_rho}). Before presenting these data, we first comment on two other
perspectives of the model.

Unlike TASEP, AEP is not particle-hole symmetric. The holes'
perspective proves rather useful. When a hole at site $i$ is chosen and can move
to site $i-1$, it ``pulls along'' the
 \textit{next} hole, provided the latter lies within $\left[ i+2,i+1+\ell _{\max }
\right] $. A picturesque way to regard this action is that of a train, a language to which we will return extensively. A third alternative description
is the zero range process (ZRP) \cite{EvansHanney05}, to which TASEP can
be mapped. On a 1-D ring of sites labeled by $\alpha =1,...,H$, each can be
occupied by an unlimited number of particles, $\ell _{\alpha }$, piling up in a column. A random site is chosen and the particle at the top is moved
to the next site. It is clear that the sites in ZRP represent the holes in
TASEP (with $H=L-N$), while $\ell _{\alpha }$ represents the cluster of
particles behind hole $\alpha $. The modification for AEP is simple: A
particle at site $\alpha $ moves by an extra step (i.e., to $\alpha +2$)
provided $\ell _{\alpha +1}\in \left[ 1,\ell _{\max }\right] $. 

\textit{Simulations and Analytic Understanding}: Deferring a more systematic investigation of AEP to another publication \cite{DKZ}, we report the highlights of our findings, mainly for a $L=1000$ ring with a
range of $\rho $ and $\ell _{\max }$. The average current $J$ is measured by the total moved particles per Monte Carlo step(MCS) over $10^6$ MCS. Fig.\ref{fig:J_rho}(a) shows that $%
J_\text{AEP}\left( \rho \right) $ is indeed $\geq \rho \left( 1-\rho \right) $
everywhere and displays more complex behavior. For $\ell _{\max }\lesssim 10$
or $\ell _{\max }=L$ , it is a relatively smooth function, much like $J_\text{TASEP}$. However, for $10\lesssim \ell _{\max
}\lesssim 900$, there is a sharp transition to a lower current branch. More
remarkably, this regime is well described by $J=1-\rho $ i.e., the holes hop as if they are non-interacting at unit velocity! In summary, $J_\text{AEP}$ follows one of two $\ell _{\max }$%
-independent functions: $J_\text{AC}\left( \rho \right) $ or $J_\text{UV}=1-\rho $.
The jump from one to the other is quite sharp. Located at some $\rho
_{c}\left( \ell _{\max }\right) $, it appears to approach a discontinuity
singularity in the thermodynamic limit. As the fuzzy lines in Fig.\ref%
{fig:J_rho} imply, large fluctuations are associated with the transition
region, the details of which remain to be systematically studied. 

\vspace{-0.4cm}

 \textit{AC Branch}: Given that two particles can move in the
same attempt, the roughest estimate for $J_{\text{AEP}}$ would be $2J_\text{TASEP}$. Indeed, $J_\text{AC}$ is qualitatively so, yet subtly different, as shown in Fig.\ref{fig:J_rho} (a). Similar to the $\lambda <1/2$ systems in \cite{GR11}, $J_\text{AC}\left( \rho \right) $ also has a region of truly accelerated or
``facilitated'' motion, i.e., $d^{2}J/d\rho
^{2}>0$. Unlike those systems,  $J_\text{AC}$ is much larger, while the
facilitated region is limited to $\rho \lesssim 0.2$. Other notable features
of $J_\text{AC}$ include: Its maximum occurs at a density beyond $1/2$; it
exceeds $2J_\text{TASEP}$ for most of the $\rho >1/2$ region; and the inflection
point appears to coincide with the maximum of $2J_\text{TASEP}-J_\text{AC}$.
Understanding these features remains a challenge \cite{DKZ}.

\emph{``Hole-train'' in UV}: To gain some insight on the striking UV behavior, we begin with a simpler system: An infinite lattice completely filled except for $H$ holes and $\ell _{\max
}\rightarrow\infty$. After presenting the exact solutions for 
$H=1,2,3$, we provide a general picture of 
``entrainment.'' In this scenario, on average, the holes
are loosely bound to some finite length and move together with UV. This
picture naturally evokes the term ``a hole-train.''

Clearly, a single hole moves with UV. With two holes, let $\ell $ denote the
gap between them. If the first hole (on the left in our model) is chosen
for update, only the first moves if $\ell =0$, while both holes move if $%
\ell >0$. If the second is chosen, the first stays and $\ell $ decreases by
unity (if $\ell >0$). Since a positive $\ell $ can never increase, the
system ends up in an ``absorbing state''
with $\ell =0$ or $1$. In other words, the two holes form a \textit{tightly bound}
state, with equal probability to be in either $\ell $, while the average
velocity is easily found to be unity \cite{DKZ}. We will refer to such a
pair as an ``engine'' (of a train). The
first non-trivial case is $H=3$. Since the third hole has no effect on the
first pair, an engine will eventually form and proceed with UV. The third
hole, naturally named the ``caboose,'' can trail the engine by any distance, $m$. Thus, a
configuration is uniquely specified by $\ell $ (0 or 1, as above) and $m$,
and a complete solution is given in terms of the probabilities $p\left(
\ell ,m\right) $. Writing a master equation for $p$ and following standard
generating function techniques, we find that, apart from the first few
terms, the stationary $p$'s decays as $\zeta ^{m}$, with $\zeta =2-\sqrt{2}$%
, so that $\left\langle m\right\rangle \thicksim \zeta /\left( 1-\zeta
\right) =\sqrt{2}$ \cite{DKZ}. In
other words, the caboose is also \textit{bound}, though not as tightly as
the first two holes. It trails with an exponential tail of characteristic
spacing $\mu =-1/\ln \zeta \cong 1.85$. We can show rigorously that this
3-hole-train moves with UV on average. Proceeding to $H>3$, we can
provide the following convincing argument (as an exact solution is being
sought) for entrainment. The total train length -- from the first hole to
the last -- can change only when three particular holes are chosen for
update: The first, the last, and the next-to-the last. Choosing the
first always increases the length by unity. If one of the other two is
chosen, the length may either remain unchanged or \textit{decrease} by
unity. Na\"{\i}vely then, for every chance to lengthen the train, there are
two which may shorten it! This length cannot fall below $H$ and
so, the expected exponential tail also emerges! Moreover,
as the engine necessarily moves with UV, the whole train must also move at
UV. In the ZRP framework, this picture is even easier to grasp. The system
corresponds to an open lattice of $H$\thinspace sites, with \textit{unit}
entry rate -- regardless of the contents of the first site. At the opposite
end, the exit rate may be as high as two -- when the last two sites are
both occupied \textit{and} chosen for update. If the filling fraction happens to
be high, the imbalance of entry-exit rates will bring it down. Thus, the
overall density is expected to remain finite. Meanwhile, being limited by
the entry rate, the average current is precisely unity.

Next, we turn to applying these results to our ring where $\ell _{\max
},L<\infty $. While UV for $H=1$ is trivial, the non-trivial role of $\ell
_{\max }$ already emerges when $H=2$. If $\ell _{\max }\geq L$, every
attempt results in both holes moving (except for $\ell =0$, which
quickly becomes $\ell =1$). Indeed, $\ell $ never changes and we have $4$
moves in one MCS, regardless of which particle is chosen. Thus, we have many
``absorbing states,'' each with current ($4/L$) that even exceeds 
$2J_{\text{TASEP}}=2\left( 2/L\right) \left(
1-1/\left( L-1\right) \right) $! On the other hand, for small $\ell _{\max }$ (say, $10$),  two holes
far apart will perform independent (totally biased) random walks so that
there are $2$ moves per MCS. However, fluctuations will cause the
smaller of the two gaps between them to fall below $\ell _{\max }$. From
this point on, $\ell $ cannot increase, as the lead hole will always pull
the trailing one. This scenario continues until the pair forms an engine moving
with UV. Here, the finiteness of $L$ is irrelevant.

\begin{figure}[bph]
\vspace{-0.5cm}
\begin{center}
\includegraphics[height=4.5cm,width=7.5cm]{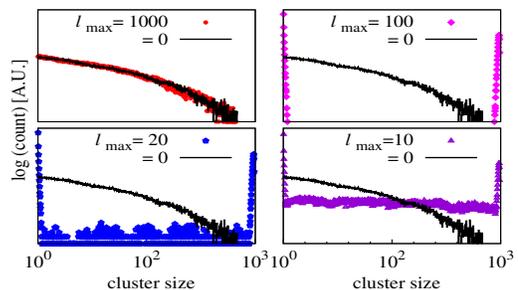}
\end{center}
\vspace{-1.1cm}
\caption{Cluster-size distribution for $H=5$ in $L=1000$. Broad distributions for the ordinary TASEP and AC are essentially identical. In phase-segregated UV, dominant maxima prevail at both extremes. For quantitative comparisons, the TASEP distribution (black line) is shown in all panels.}
\label{fig:cluster}
\end{figure}

For $H=3$, the role \textit{and value} of $\ell _{\max }$ becomes more
significant, since the caboose is \textit{loosely }bound and $\mu $ must
enter somewhere. In particular, if $\ell _{\max }$ is too small, the caboose
can easily come unbound, wanders around the ring and then ``unbinds'' the engine by pulling the first hole away from
the second. On the other hand, if $\ell _{\max }$ is too large (e.g., $%
O\left( L\right) $), the caboose directly affects the engine's
integrity. A clear picture now emerges: For $\mu \ll \ell _{\max }\ll L$,
the 3 holes tend to be entrained and move with UV. Indeed, simulations with 
$\ell _{\max }=20$ shows that, in 4000 measurements, the train length never
exceeds $15$, a fact entirely consistent with $\zeta^{15}\thicksim 10^{-4}$. This picture extends easily to $3<H\ll L$ as
 a hole-train moving at UV for a range of $\ell _{\max }$. To provide a more
quantitative view of the role played by $\ell _{\max }$, we show various cluster
size distributions in the $H=5$ case in Fig.\ref{fig:cluster}. With $\ell _{\max }=0$ (ordinary TASEP), we find a
broad distribution of sizes, implying the absence of entrainment. The
average cluster size, as expected, is $\thicksim200=L/H$. At the other extreme ($%
\ell _{\max }=1000$), we find a statistically indistinguishable
distribution! This result is also understandable, especially from a ZRP
perspective, where particles just move around the five sites, at typically
twice the speed. The other three distributions are drastically different,
showing dominant peaks at both ends, a marked signal of strong clustering of
the 5 holes. For $\ell _{\max }=10$ and $20$, the system is clearly
attempting a transition to the entrained state. There is a small but broad
distribution of sizes in between, implying that one of the holes becomes
unbound, creating intermediate size gaps as it wanders around the rest of
the ring. By $\ell _{\max }=100$, such events do not occur in our runs.
Indeed, the frequency of the large cluster shows that just one such
cluster appears in each measurement. Of course, its size is precisely the
complement of the length of the hole-train. In this case, the latter is $\thicksim
8.92$, approximately $2H-1$. In short, the hole-train has, on average, one spacing between its cars. We will show in a more systematic study that this result prevails in the UV phase for a large range of $H$\cite{DKZ}.
\vspace{-0.4cm}

Given the existence of a UV phase, let us turn briefly to the transition to
the AC phase. From the viewpoint of the hole-train, it is an
``unbinding'' transition. As more holes are
added to a finite ring, the size of the macroscopic cluster separating the
engine from the caboose, $\lambda $, decreases. When it drops to $\thicksim \ell _{\max }$, the caboose
will ``pull the engine apart.'' Meanwhile,
if we assume the train length to be $\thicksim2H$, we have $\lambda
\thickapprox L-2H$, leading to a rough estimate $\rho _{c}\thicksim \left(
1+\ell _{\max }/L\right) /2$. Remarkably, this estimate is within 10\% of the data points gathered so far \cite{DKZ}. In the transition region, the
train dissolves and re-forms, resulting in the large fluctuations we observe
(in Fig.\ref{fig:J_rho}). To find a good estimate for $\rho _{c}$ is
non-trivial since these fluctuations will undoubtedly play important roles. 

\textit{Concluding Remarks}: Inspired by long-range interactions among particles such as the speed-up in transcription through cooperative RNAPs, we investigate AEP in which a particle
hopping onto a cluster of length up to $\ell _{\max }$ simultaneously triggers
the first particle in that cluster to hop. This extension from the paradigmatic TASEP gives rise to various novel properties such as the transition from homogeneous to phase-segregated and the intriguing unit-velocity phase. Simulating such a system on a $L=1000$ ring with various filling fractions $\rho $, we find the augmentation of $J\left( \rho \right) $ over the ordinary $J_\text{TASEP}=\rho \left(
1-\rho \right) $. Surprisingly, we discover that when $\rho $ exceeds a
critical $\rho _{c}\left( \ell _{\max }\right) $, a 
``condensation'' transition takes place and the system
becomes phase segregated. Here the holes gather into a loosely bound
cluster, moving as a whole around the ring, motivating us to name it a
``hole-train.'' Even more remarkably, this train moves with \textit{unit
velocity}! Focusing on the UV phase, we measured cluster-size
distributions and sought theoretical understanding. Our studies with small number of holes provided
adequate insight for us to understand why entrainment exists, why the train
should move at UV, and how the transition from the homogeneous AC phase
arises. The role of $\ell _{\max }$ is critical: If it is too small,
binding cannot be sustained; too large, the two ends of the train
interact and the engine/train disintegrates. Perhaps these insights will
help us arrive at a full analytic theory. 

Many other intriguing issues remain, on both theoretical and modeling fronts. We should first emphasize that, at this stage, AEP should be viewed more as a significant extension of TASEP than an explicit model for transcription. To address the former, the most immediate need is a systematic 
study with a range of $L$'s so that finite size effects can be quantified. Locating the transition and providing quantitative characterizations (e.g., fluctuations or full distributions of $J$'s) will be revealing. This can also expose the nature of the transition. Does it display the same behavior as a typical first-order transition? If so, does its hysteresis follow standard properties? Does $\rho _{c}\left( \ell _{\max },L\right) $
scale to $\rho _{c}\left( \ell _{\max }/L\right) $? What correlations, spatial and temporal, can we expect? Structure factors, cluster-size distributions, and power spectra for various quantities should be studied to provide us a more complete picture. Moreover, is there a better estimate of the average train length ($2H-1$) than the rough values from $\mu $ or $\left\langle m\right\rangle $? Of course, interesting questions also abound for the AC phase (e.g., a system with $\ell _{\max }=L$). As $L\rightarrow \infty $, does $J_\text{AC}$ attain its maximum at $\rho >1/2$, as our data seem
to indicate? If so, how can we predict the values of both $\rho $ and $J_\text{AC} $? Similarly, can we understand the significance of the
inflection point and where is it located? The ultimate goal is to find the exact steady state distribution
$P\left( \left\{ n_{i}\right\} \right) $, likely a highly non-trivial task. Furthermore, we can raise all the intricate questions which were directed at the ordinary TASEP, e.g., dynamic properties and large deviation functionals. Turning to the modeling front, we should move beyond the simple AEP and consider more complex rules for ``pushing'' in real applications. Rules that readily come to mind include cluster-size dependent triggered moves, open AEP, extended particles, inhomogeneous hopping rates, multiple particle species, etc\cite{ZDS2011}. We hope that AEP will open a new chapter for exclusion processes as well as a new window into non-equilibrium statistical mechanics in general.

\section*{Acknowledgements}
We thank T. Chou, R.J. Harris, K. Mallick, S. Redner and B. Schmittmann for illuminating discussions.
JJD and RKPZ are grateful to the hospitality of MPIPKS, Dresden, where some of this research was carried out.
This research is supported in part by grants from the US NSF DMR-1104820 and DMR-1005417.

\bibliographystyle{apsrev4-1}

\bibliography{references_aep}

\end{document}